\RequirePackage{fix-cm}
\documentclass[smallextended]{svjour3}       % onecolumn (second format)

\smartqed  % flush right qed marks, e.g. at end of proof
\usepackage{graphicx}
\usepackage{braket} %edited%
\usepackage{physics} %edited%
\usepackage{amsmath} %edited%
\usepackage{mathtools} %edited%
\graphicspath{ {images/} } %edited%
\usepackage{color}

\begin{document}

\title{On Interchangeability of Probe--Object Roles in Quantum--Quantum Interaction-Free Measurement 
%\\ or \\
%Quantum Interaction--Free Measurement Under Assumption of Symmetry Between Object and Probe %\thanks{Grants or other notes
}


\author{Stanislav Filatov\and Marcis Auzinsh}

\authorrunning{Short form of author list} % if too long for running head

\institute{Stanislav Filatov \at
              Department of Physics,
 University of Latvia, Raina boulevard 19, 
 LV-1586, Riga, Latvia \\        
           \email{sfilatovs@gmail.com}            \\
          \and
           Marcis Auzinsh \at
              Department of Physics,
 University of Latvia, Raina boulevard 19, 
 LV-1586, Riga, Latvia \\        
           \email{Marcis.Auzins@lu.lv}   
}

\date{Received: date / Accepted: date}

\maketitle

\begin{abstract} 

In this paper we examine Interaction-free measurement (IFM) where both the probe and the object are quantum particles. We argue that in this case the description of the measurement procedure must by symmetrical with respect to interchange of the roles of probe and object. A thought experiment is being suggested that helps to determine what does and what doesn't happen to the state of the particles in such a setup. It seems that unlike the case of classical object, here the state of both the probe and the object must change. A possible explanation of this might be that the probe and the object form an entangled pair as a result of non-interaction.
\end{abstract}
\section{Introduction}
\label{intro}
The quantum measurement problem from the very beginning of quantum mechanics has been causing many discussions and was one of the stumbling blocks for the interpretation of the mathematical formalism of the theory \cite{Jacobs2014}. It will not be an exaggeration to say that the question how the wave-function of the object, that is being measured collapses, never was convincingly answered.

In the beginning of 1950-ies Mauritius Renninger widened the scope of the discussion about measurement in quantum mechanics, when he asked the questions -- whether the fact that there was a non-zero probability for the object to interact with the measuring device, but this interaction did not occur, still can be regarded as a measurement -- in this case interaction-free measurement (IFM)\cite{Ren1953}?

This discussion was taken to the new level by Elitzur and Vaidman \cite{EV1993} when they proposed a famous thought experiment about the bomb, that is placed in one of the arms of Mach--Zehnder interferometer and a single photon is passing through this interferometer. In this case we have a quantum probe (photon) which, when faced with the first beam splitter has equal probability to go through the semitransparent mirror or to be reflected. One of its paths inside of the interferometer can be potentially blocked by a large classical object that gets probed. In case the object is present, it can be revealed by the detector click in one of the output ports after the second beam splitter of the interferometer which otherwise due to destructive interference of the two photon pass-ways, never clicks. \textcolor{blue}{Further on we will call such a setup Elitzur-Vaidman Interaction-Free Measurement (EV IFM).}

Such protocols have been implemented experimentally and their efficiency boosted to the maximum by using quantum Zeno effect. (Zeilinger, Kwiat, Kasevich et.al. \cite{Kwiat1995}). 

It has been suggested already in the pioneering paper by Elitzur and Vaidman \cite{EV1993} and then by Kwiat et. al. \cite{Ref3} that an object needs not be a classical non-transparent object. It might also be a quantum object in a superposition state of ``being on the way and not being on the way". However, those early attempts to analyze quantum object haven't been pursued  with the same vigour as the case of classical object. 

New aspects of the IFM, that often are even called paradoxical, were analyzed by Lucien Hardy \cite{Ref4} who suggested to examine two Mach-Zehnder type interferometers but for particles. In one of them an electron travels in another one -- positron. At some point in space the path of the the electron in one interferometer overlaps with the path of the positron in another interferometer. 

The paradox arises from the fact that, in contrast to the initial experiment proposed by Elitzur and Vaidman,  the path for the electron and the positron gets blocked only when both particles are taking the spatially overlapping arms of interferometer and annihilate. This makes it impossible to block path only potentially when particle is not taking that path. Hardy's paper initiated a series of papers (see for example \cite{Ref10,Ref11,Ref12,Ref13}) that deal with the paradox.

A new revival of the idea of IFM with a quantum object was triggered by the realization of its usefulness for quantum computation and information processing. So Azuma \cite{Ref5} suggested to use IFM to construct logical quantum gates. Some researchers started to consider possibilities of creating logical gates based on light-atom interaction (or more precisely, non-interaction) using the internal degrees of freedom of atoms (Moore et.al. \cite{Ref6}). Ideas on utilizing principles of IFM for quantum computation \cite{Ref14,Ref15,Ref16} quantum communication \cite{Ref17,Ref18} and quantum cryptography \cite{Ref19} have spurred ever since. And finally  IFM approach was used to analyze interaction between distant atoms \cite{Ref20}.

These papers lead to more and more detailed analysis of the interaction free measurements of such quantum objects as an atom with internal degrees of freedom by a polarized light. So Potting et al. \cite{Ref7} explored to what extent IFM is non-perturbative. They utilized for that analysis superpositional internal degrees of freedom of atom and photon.  In that paper the idea that sometimes, as a result of non-interaction, entanglement between atom and light might occur is implicitly introduced. A paper based on it \cite{Ref8}, is introducing explicitly entanglement between internal degrees of freedom of a quantum object and photon's polarization, however, only for some cases of non-interaction that result from specific initial conditions of the system. In \cite{Ref9} entanglement is shown to lie conceptually at the heart of Elitzur-Vaidman original IFM proposal. 

In short, there are two approaches to quantum-quantum non-interaction. First is to carry over the assumption of the classical IFM, namely that the object is ``rigid'' or that its state is unaltered by non-interaction and the state of the probe accommodates all of the change if there is any. Second approach is to allow both the state of the object and the probe to change as a result of non-interaction, such treatment has lead some researchers towards introduction of entanglement as a result of non-interaction in specific configurations on probe-object system \cite{Ref7,Ref8,Ref9}. 

In this paper we demonstrate that only the second approach is consistent with rigorous treatment of IFM. We require that the description of non-interaction is invariant if we exchange the roles of ``probe'' and ``object'' between the components of the IFM. This is done by means of a thought experiment by which any potential description of non-interaction can be tested for self-consistency. We show that none of the previously proposed descriptions (including those involving entanglement) pass the aforementioned test. We propose a  description where entanglement is introduced in {\em all} cases of non-interaction, regardless of the initial conditions of the system, which does pass the test. The aforementioned thought experiment and the new candidate descriptions are the contributions of this paper.

\textcolor{red}{In parallel to Elitzur and Vaidman approach to IFM there are several other methods and thought experiments proposed to implement interaction free measurement. For example, very interesting method was suggested by Marlan Scully, Gerthold-George Englert and Herbert Walther \cite{Ref21}. Their idea assumes that two micromasers are inserted in each path of atoms traveling in a double slit type interferometer. This allows with infinitesimally week interaction of atom with cavity to determine which pass in the interferometer atom has traveled.} 

\textcolor{red}{Others use the term "interaction-free" for experiments where nothing gets probed, but some information is revealed about the path of the photon wavepacket in seemingly interaction-free manner \cite{Ref22}.} 

\textcolor{red}{Finally, Elitzur Vaidman IFM method can be considered as an example of a broad class of quantum non-demolition measurements considered by Braginsky et.al \cite{Brag80}  that leaves the quantum state of the object undisturbed provided it was in an eigenstate prior to the measurement.}

\textcolor{red}{Partially interaction free measurement can be seen in the context of the extended measurement theory \cite{Putz2010} and there are parallels with entanglement at the chemical bond level \cite{Putz2010bond}.}

\textcolor{red}{
A great survey of the subtleties of "Interaction-free" in the Elizur Vaiman (EV) sense is given by Lev Vaidman (see \cite{Ref23} and references therein). It is shown there, for example, that EV IFM is different from general null-result experiment because it doesn't require the knowledge of the object's wavefunction prior to the experiment and that EV IFM is not absolutely energy- or momentum-exchange free. Also, although EV IFM allows for detection of the classical object without any particles travelling in its vicinity, this claim is only valid to an extent if the object is quantum. We adopt EV IFM definition with, as will be seen from the rest of the paper, with the modification that quantum probe and object are interchangeable in their roles and that quantum object may occupy interferometer arms superpositionally. All the limitations of EV IFM, especially as our object is quantum, hold for our setup as well.}

The rest of the paper is structured as follows. 

Part \ref{model} introduces all the machiery used in the following thought experiment, discusses the terms and main assumptions behind them. It also contains an analysis that reveals an initial intuition behind this work. Part \ref{hermetic} presents a thought experiment that establishes restrictions on successful description of non-interaction. It is the heart of the work where the main argument is presented. A very insightful reader might just examine the figure \ref{fig:filter} that is a condensed thesis of this paper. Part \ref{conclusion} is conclusion.

\section{The model for IFM of quantum object}
\label{model}

\begin{figure}[h!]
    \centering
    \includegraphics[width=0.3\textwidth]{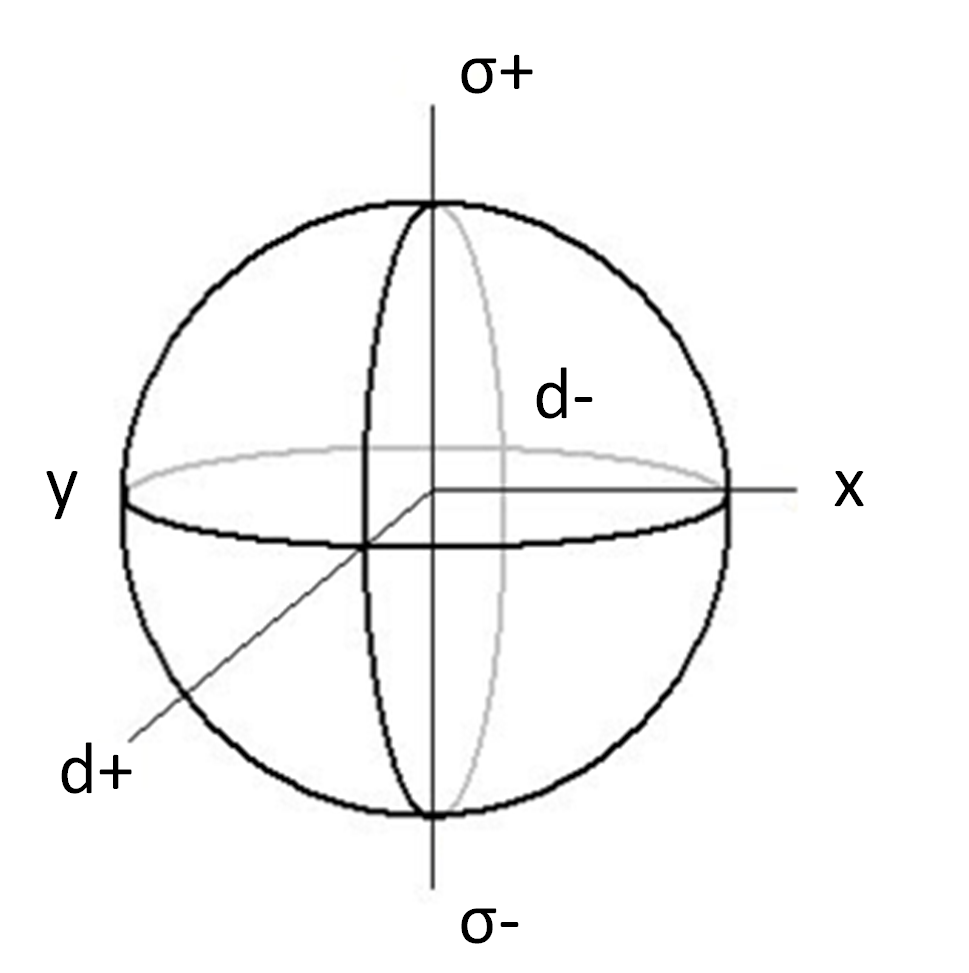}
    \caption{Bloch sphere representation of a two-level quantum system. X and Y stand for horizontal and vertical polarization of photon or atom, $\sigma+$ and $\sigma-$ for left- and right-hand circular, d+ and d- for ``diagonal" polarization. In general, the quantum object need not be a photon or an atom. Particle's spatial location after passing through 50/50 beamsplitter can be represented by the same Bloch sphere. Note that orthogonal states like $\ket{\sigma+}$ and $\ket{\sigma-}$ are facing 180 degrees away from each other}
    \label{fig:blochsphere}
\end{figure}

As it was shown in the Introduction, there are different models of quantum object and probe in IFM. In order to avoid ambiguity, we describe what in this paper will be understood by the quantum probe and the quantum object. Generally, both object and probe are two-level quantum systems (see fig. \ref{fig:blochsphere} for a Bloch-sphere representation of a two-level quantum system). Their quantum states are linked to each other by potential interaction. Probe is used to inspect the state of the object. The two states can be both internal (polarization of light or energy eigenstates of an atom) and external (path taken inside an interferometer) degrees of freedom. This analysis employs internal degrees of freedom of both object and the probe.

\begin{figure}[h]
    \centering
    \includegraphics[width=0.5\textwidth]{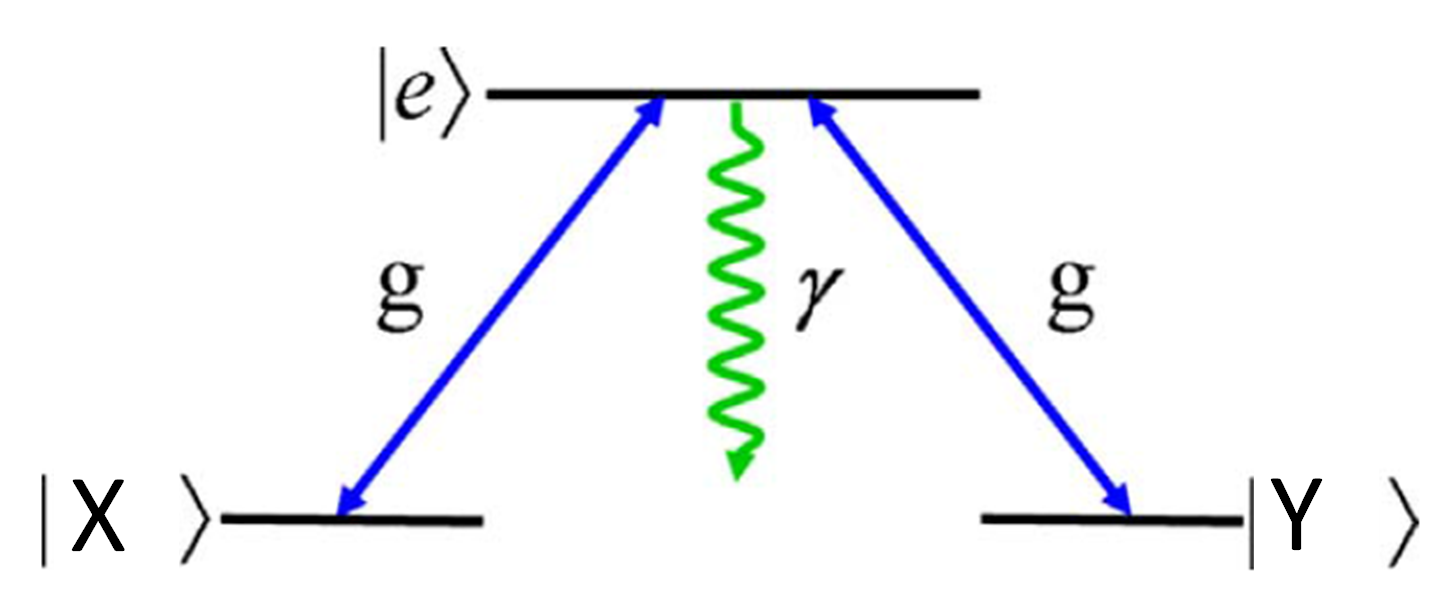}
    \caption{Two-level scheme of Moore et. al \cite{Ref6}. Atom has two ground states $\ket{x}_{at}$ and $\ket{y}_{at}$ that are coupled to the excited state $\ket{e}$ through interaction with a photon of state $\ket{x}_{ph}$ or $\ket{y}_{ph}$ respectively. $\ket{e}$ is an excited state that immediately decays ($\gamma$). 
    }
    \label{fig:two-level}
\end{figure}

{\textbf{Quantum object}} 
In this analysis is a two-level atom with two orthogonal quantum states (see fig. \ref{fig:two-level}). When IFM will be conducted, the object will initially be in a coherent superposition of those two states. Each of these two states can potentially interact with specific polarization component of an incoming photon (probe). To analyze this process we are forced to introduce auxiliary rapidly decaying into the thermal bath excited state of the atom. Rapid decay, according to \cite{Ref6}, is very important for this model. First, it ensures that if the interaction has happened the measuring photon disappears into the thermal bath and there is no possibility of reabsorbtion of the newly emitted photon. Second, the fast decay implies that the ``thermal bath" is constantly measuring whether the atom is in the excited $\ket{e}$ state so we can be sure if we don't see any emission, the atom is in the ground state.

As a {\textbf{Quantum probe}} 
in this analysis will serve a photon with two states of polarization as internal degrees of freedom. We examine superpositions of internal states of polarization because for those there is no intuitively appealing basis-of choice, unlike for spatial degrees of freedom where the basis of preference is the one with apparently distinct channels through which the particle can travel, as in the case of Mach-Zahnder interferometer.

{\textbf{Non-interaction}} 
\textcolor{red}{Formally Quantum measurement as an interaction as well as a non-interaction can be analyzed by means of projection operator, often called von Neumann projection operator \cite{Neu2018}. However, in this paper we decided to use more informal approach, in our opinion, allowing to develop and employ quantum intuition, leaving more formal analysis for further studies.}

We consider atomic and photonic states (or degrees of freedom) to be ``aligned" when the probability of interaction is maximally possible. Furthermore, we assume idealized photon and atom, so when their states are aligned, interaction probability is 100\%. Such idealization makes clear what is meant by ``non-interaction". It is a result of state misalignment, not the result of imperfect match of trajectories or a mismatch between the frequency of the photon and that of the transition. Still though, analysis presented here holds even if we allow for distortions, imperfections and noise (a short case study is given at the end of Part \ref{hermetic} .

\begin{figure}[h!]
    \centering
    \includegraphics[width=0.6\textwidth]{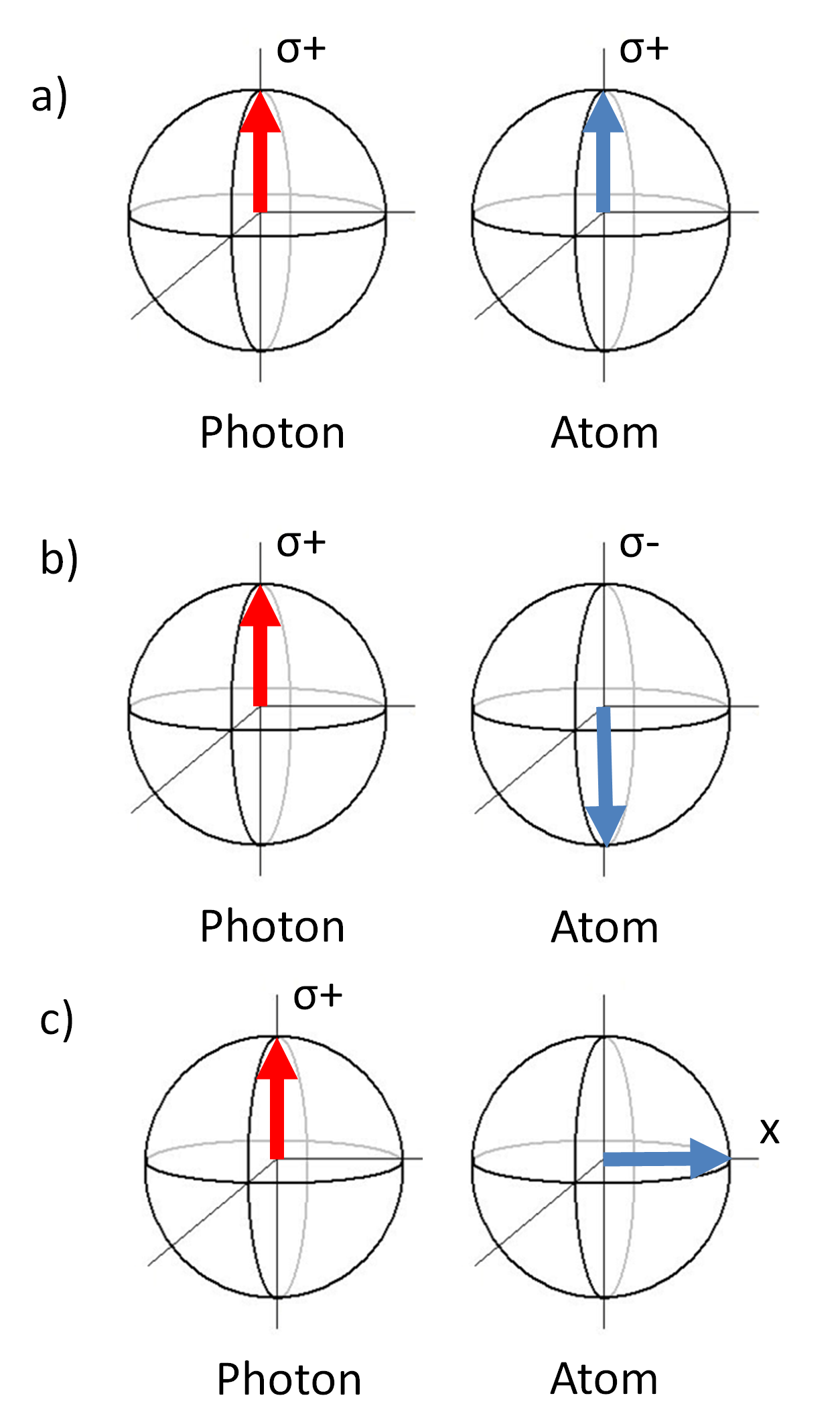}
    \caption{Bloch sphere representation of states that a) will always interact - in this case it's state $\ket{\sigma+}_{ph}\ket{\sigma+}_{at}$, b) will always ``non-interact" - in this case $\ket{\sigma+}_{ph}\ket{\sigma-}_{at}$ 
    and c) will interact only in half the cases  $\ket{\sigma+}_{ph}\ket{x}_{at}$ }
    \label{fig:al-un}
\end{figure}

The aforementioned proposition can be graphically represented by a set of Bloch spheres (see fig \ref{fig:al-un}). Photon and atom which always interact are depicted by a pair of vectors that are pointing in the same direction on the Bloch sphere (see fig \ref{fig:al-un} a.). And photon and atom which have zero probability of interaction are represented by a pair of vectors on the Bloch spheres that are pointing in the opposite directions (see \ref{fig:al-un} b.). Note that vectors that are Hilbert-space orthogonal are represented on the Bloch Sphere as pointing 180 degrees apart. Every arbitrary pure atomic state has one corresponding photonic state that will result in interaction in 100\% of cases and one (orthogonal to the previous) which will result in interaction in 0\% of the cases (this is shown more formally in Appendix).

In this work we exploit a property of quantum mechanics that interactions are basis-of-description independent. We will be examining two bases that are related to each other as

\begin{equation} \label{basischange}
\begin{split}
\ket{\sigma+}&= \quad (\ket{x} + i\ket{y})/\sqrt{2} \\
\ket{\sigma-}&= \quad (\ket{x} - i\ket{y})/\sqrt{2} \\
\ket{ \ x \ }&= \quad (\ket{\sigma+} + \ket{\sigma-})/\sqrt{2} \\
\ket{ \ y \ }&= -i(\ket{\sigma+} - \ket{\sigma-})/\sqrt{2}
\end{split}
\end{equation}

So if we chose (x ; y ) basis instead of $(\sigma+ ; \sigma-)$ basis for both photon polarization and atomic state, the atomic state  $\sigma+$ or $(x+iy)/\sqrt{2}$ should always get excited by the corresponding $\sigma+$ polarization state of the photon. In other words, there is no ``preferred" set of basis vectors in which to represent the state of the atom or the photon (see figure \ref{fig:blochsphere} for a pictorial representation of possible qubit states). 

We will denote the state of photon-atom system in the following way:
$\ket{\phi}_{ph} \ket{\omega}_{at}$ where the state of photon is in the first ket vector, and the atomic state is in the second ket vector. The subscripts, although being redundant, are there to remind the reader the abovementioned arrangement. 

{\textbf{IFM in case of partially overlapping states}}

Let us look at the partially overlapping states in some detail. As an example let us use the state depicted in the fig. \ref{fig:al-un} c).

Before the possibility of interaction the joint state of atom and photon is
\begin{equation} \label{e01}
\begin{split}
\ket{\sigma+}_{ph}\ket{x}_{at}
\end{split}
\end{equation}
In order to examine interaction/non-interaction we need to rewrite both states in the same basis. Two most obvious choices would be to use either eigenbasis of the atom or the eigenbasis of the photon. Using the aforementioned relation of the bases (eq. \ref{basischange}), if we  are using atomic eigenstates, we arrive at 
\begin{equation} \label{e001}
\begin{split}
\ket{\sigma+}_{ph}\ket{x}_{at} =
 (\ket{x}_{ph} \ket{x}_{at} + i\ket{y}_{ph} \ket{x}_{at})/\sqrt{2}
\end{split}
\end{equation}
This means that if interaction happens, the $\ket{x}_{at}$ state of the atom absorbs the $\ket{x}_{ph}$ state of the photon; photon energy is dissipated, photon disappears and the atom decays to some unknown state in thermal bath. But there is 50\% chance of non-interaction. In this case the atom remains in the initial state $\ket{x}_{at}$ and photon appears in the $\ket{y}_{ph}$ state.

Equally well we can use the eigenbasis of photon as the basis of description. Then similarly, the joint state of atom and photon will be
\begin{equation} \label{e1}
\begin{split}
\ket{\sigma+}_{ph}\ket{x}_{at} = (\ket{\sigma+}_{ph}\ket{\sigma+}_{at} + \ket{\sigma+}_{ph} \ket{\sigma-}_{at})/\sqrt{2}  
\end{split}
\end{equation}

In this case, non-interaction would mean that photon remains in $\ket{\sigma+}_{ph}$ state and the atom appears in $\ket{\sigma-}_{at}$. In the first case it looks like  the state of the atom is ``rigid" and the sate of the photon) is ``fluid", meaning that the state of the atom doesn't change during the non-interaction, but the state of the photon accommodates in such a way to be orthogonal to the state of the atom. This resembles the most the case of the classical IFM when the object (opaque classical massive obstacle on the way of the photon) obviously doesn't change during the non-interaction. In the second case, the situation is directly opposite - photon appears to be ``rigid", but atom - ``fluid" in the abovementioned sense.

Existence of these two possible approaches seems to create a problem. As it is not clear to which basis set the preference should be given as both the object and the probe are similar two-state quantum systems and are representable by Bloch spheres and should have symmetrical properties. In the next section we provide an argument why none of the particles can remain unchanged. In what follows a thought experiment is presented which shows that the state of both particles must change in the event of non-interaction. Even more, after non-interaction the states of quantum object and probe are not separable anymore, in other words they become entangled.  

\section{The hermetic argument}
\label{hermetic}

\begin{figure}[h!]
    \centering
    \includegraphics[width=0.5\textwidth]{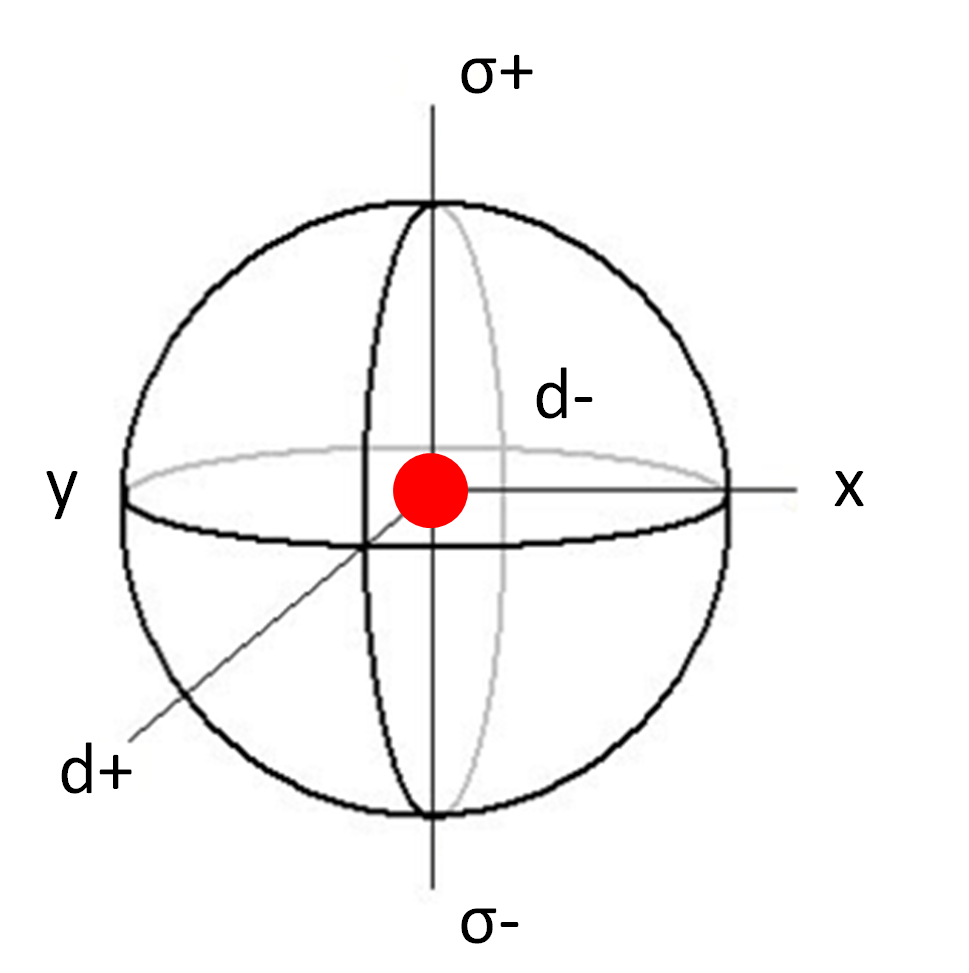}
    \caption{Bloch sphere representation of the Identity density matrix. A dot at the center of the Bloch sphere.}
    \label{fig:densitymatrix}
\end{figure}

To prove the abovementioned claim, let us consider two set of photons. In first set, photons are with equal probability either in bases state $\ket{x}_{ph}$ or in bases state $\ket{y}_{ph}$. In the second set photons are prepared with equal probability either in a bases state $\ket{\sigma+}_{ph}=(\ket{x}_{ph}+i\ket{y}_{ph})/\sqrt{2}$ or in a bases state $\ket{\sigma-}_{ph}=(\ket{x}_{ph}-i\ket{y}_{ph})/\sqrt{2}$. It's known that both sets are experimentally indistinguishable and sets are represented by the same density matrix (see equations \ref{eq:rho1} and \ref{eq:rho2} and fig. \ref{fig:densitymatrix} for a Bloch-sphere representation) \cite{Blum}. 

By the means of the following thought experiment we will show that any description of non-interaction that assumes one of the partners of the IFM to be more ``rigid" than another, is incompatible with the density matrix indistinguishability condition just stated.

\begin{equation}
\begin{split}
    \rho_1 = &\frac{1}{2}(\ket{x}\bra{x} + \ket{y}\bra{y}) = \\
   &\frac{1}{2}\left(
    \begin{bmatrix} 1&\ 0\\ 0&\ 0 \end{bmatrix}
   + 
    \begin{bmatrix} 0&\ 0\\ 0&\ 1 \end{bmatrix}
    \right)
    = \\
    &\frac{1}{2}
    \begin{bmatrix} 1&\ 0\\ 0&\ 1 \end{bmatrix}
\end{split}
\label{eq:rho1}
\end{equation}

\begin{equation}
\begin{split}
    \rho_2 = &\frac{1}{2}(\ket{\sigma+}\bra{\sigma+} + \ket{\sigma-}\bra{\sigma-}) = \\
   &\frac{1}{4}\left(
    \begin{bmatrix*}[r] 1 & -i\\ i & 1 \end{bmatrix*}
   + 
    \begin{bmatrix*}[r] 1 &\  i\\ -i &\  1 \end{bmatrix*}
    \right)
    = \\
    &\frac{1}{2}
    \begin{bmatrix} 1&\ 0\\ 0&\ 1 \end{bmatrix}
    =
    \rho_1
\end{split}
\label{eq:rho2}
\end{equation}

\begin{figure}
\centering
\includegraphics[width=0.8\textwidth]{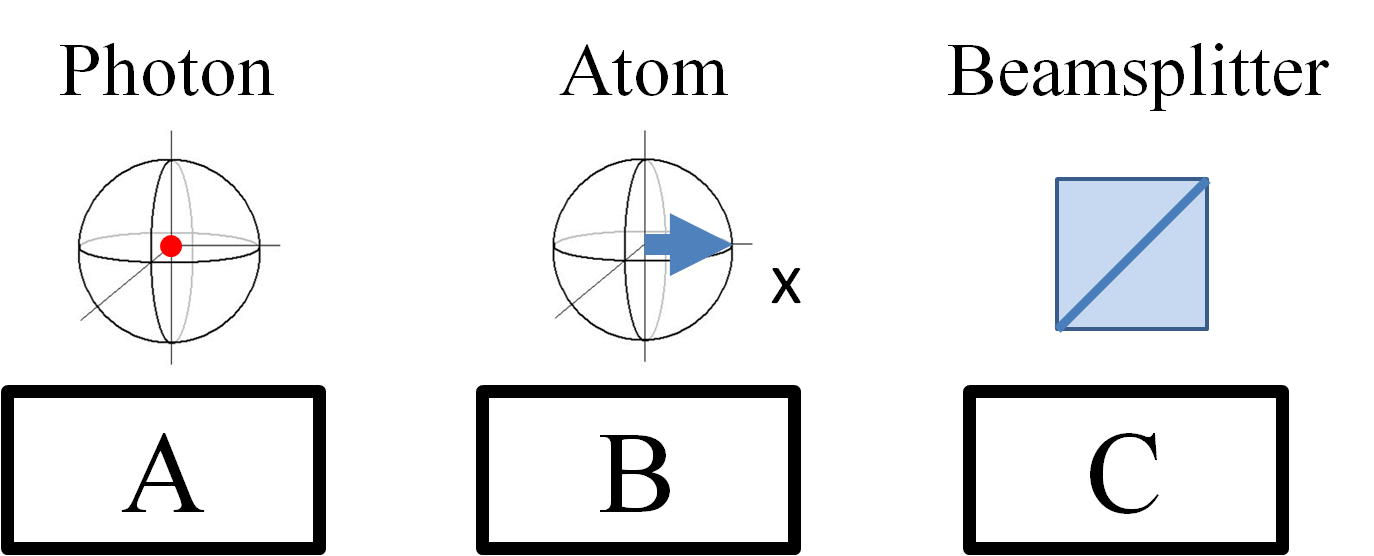}

\caption{Filter-device.  Stage A produces ``probes", in this case - photons. It operates in two modes, both of which correspond to Identity density matrix. Mode 1: photonic state is either (50\%-50\%) $\ket{x}_{ph}$ or $\ket{y}_{ph}$. Mode 2: Photonic state is either (50\%-50\%) $\ket{\sigma+}_{ph}$ or $\ket{\sigma-}_{ph}$.  Stage B is an ``object", in this case - atom. It is prepared in the state $\ket{x}_{at}$ regardless of the mode of operation of Stage A; for every photon we are using a ``fresh" atom.  Stage C is a beamsplitter that separates (projects) ``probes" (here - photons) that haven't interacted with the ``object" (here - atom) into $\ket{x}_{ph}$ and $\ket{y}_{ph}$ (basis of the ``object") which later get detected by a corresponding detector. If we assume that photonic state remains unchanged after it non-interacts with the atom, two modes of operation at the Stage A will be distinguishable. Any successful description of (non-)interaction should satisfy the restrictions proposed here also if we change the basis of the ``object" and measurement (stages B and C) as well as if we exchange the roles of ``probe" and ``object" between the photon and atom (stages A and B).}
\label{fig:filter}
\end{figure}

The thought-experimental device consists of three stages (see fig. \ref{fig:filter}) - A (Source), B (Filter) and C (Analyzer). Stage A produces ``probes", in this case - photons. It operates in two modes, both of which correspond to Identity density matrix. Mode 1: photonic state is either (50\%-50\%) $\ket{x}_{ph}$ or $\ket{y}_{ph}$. Mode 2: Photonic state is either (50\%-50\%) $\ket{\sigma+}_{ph}$ or $\ket{\sigma-}_{ph}$.  

Stage B is an ``object", in this case - an atom. It is prepared in the state $\ket{x}_{at}$ regardless of the mode of operation of Stage A; for every photon we are using a ``fresh" atom.  

Stage C is a beamsplitter that separates (projects) ``probes" (here - photons) that haven't interacted with the ``object" (here - atom) into $\ket{x}_{ph}$ and $\ket{y}_{ph}$ (eigenbasis of the ``object") which later get detected by a corresponding detector. 

Initially let us look at two opposite scenarios of non-interaction. And demonstrate that both of these assumptions lead to contradiction with the density matrix indistinguishability condition defined by the equation \ref{eq:rho1} and \ref{eq:rho2}.

Let's assume for the moment that the state of the photon remains unchanged after the non-interaction.

If the source operates in the first mode the situation would look as follows: 
Photonic state $\ket{x}_{ph}$, state of the system $\ket{x}_{ph}\ket{x}_{at}$, in physical proximity this leads to $\ket{scatter}$ (absorption of the photon with subsequent scattering into the thermal bath).
Photonic state $\ket{y}_{ph}$, state of the system $\ket{y}_{ph}\ket{x}_{at}$, in physical proximity this leads to event of non-interaction, and by assumption the state of the photon doesn't change (and we don't care what happens with the state of atom for now). So in the analyzer C we always get the resulting state of the photon $\ket{y}_{ph}$

If the source operates in the second mode, situation looks as follows:
Photonic state $(\ket{x}_{ph}+i\ket{y}_{ph})/\sqrt{2}$, state of the system $(\ket{x}_{ph}+i\ket{y}_{ph})/\sqrt{2}\ket{x}_{at}$, in physical proximity this leads to an event of scattering 50\% of the time, while with probability 50\% nothing happens and photon continues undisturbed (as we've assumed). So in the analyzer C we get resulting state $\ket{x}_{ph}$  or $\ket{y}_{ph}$  with equal probability.
Similar story with the photon in the initial state $(\ket{x}_{ph}-i\ket{y}_{ph})/\sqrt{2}$.

The first and the second modes of operation of photon source A can therefore be distinguished in this thought experiment. Although they correspond to the same photon density matrix. In the Mode 1 of operation of stage A detectors will be detecting only $\ket{y}_{ph}$ photons (only one detector ``clicks"), while in the Mode 2 of operation - both $\ket{y}_{ph}$ and $\ket{x}_{ph}$ photons (either first or the second detector ``clicks", randomly). 

Logically, the next step would be to assume that in fact during the process of non-interaction, the atom is ''rigid" and the photon - ''fluid". In the aforementioned setup it may occur that the contradiction is evaded. If the stage A operates in the first mode, producing photon with equal probability in either state $\ket{x}_{ph}$ or $\ket{y}_{ph}$, only $\ket{y}_{ph}$ photons non-interact and analyzer always sees a photon in the state $\ket{y}_{ph}$. If on the other hand, the source operates in the second mode, producing photon with equal probability in either state $\ket{\sigma+}_{ph}$ or $\ket{\sigma-}_{ph}$, in case of non-interaction (by the assumption of the rigidity of atom) both of those states get transformed into $\ket{y}_{ph}$. As a result, the detector sees only the photon in state $\ket{y}_{ph}$ again. 

At first glance it may seem that the contradiction arising from the assumption of what is ``rigid" and what is ``fluid", is avoided. But it is not. In case of non-interaction of two quantum objects the process should be symmetrical with respect to which one of the partners we call ''probe" and which ''object". In order to test this symmetry, let the source produce atoms which now will be the ''probe" and the filter will consist of photon. If the source operates in mode 1: atomic state is either (50\%-50\%) $\ket{x}_{at}$ or $\ket{y}_{at}$. Mode 2: Atomic state is either (50\%-50\%) $\ket{\sigma+}_{at}$ or $\ket{\sigma-}_{at}$. 

Stage B is an ``object", in this case - photon. It is prepared in the state $\ket{x}_{ph}$ regardless of the mode of operation of Stage A; for every atom we are using a ``fresh" photon.  Stage C is a beamsplitter that separates (projects) ``probes" (here - atoms) that haven't interacted with the ``object" (here - photon) into $\ket{x}_{at}$ and $\ket{y}_{at}$ (basis of the ``object") which later get detected by a corresponding detector.

We now use the assumption that seemed non-contradictory in the first case, namely that atom is ''rigid" and the photon is ''fluid". 
If the source operates in the first mode the situation would look as follows: 
Atomic state $\ket{x}_{at}$, state of the system $\ket{x}_{at}\ket{x}_{ph}$, in physical proximity this leads to $\ket{scatter}$ event of scattering.
Atomic state $\ket{y}_{at}$, state of the system $\ket{y}_{at}\ket{x}_{ph}$, in physical proximity this leads to event of non-interaction, and by assumption the state of the atom doesn't change (and we don't care what happens with the state of photon for now). So in the analyzer C we always get the resulting state of the atom $\ket{y}_{at}$

If the source operates in the second mode, situation looks as follows:
Atomic state $(\ket{x}_{at}+i\ket{y}_{at})/\sqrt{2}$, state of the system $\ket{x}_{ph}(\ket{x}_{at}+i\ket{y}_{at})/\sqrt{2}$, in physical proximity this leads to an event of scattering 50\% of the time, while with probability 50\% nothing happens and atom continues undisturbed (as we've assumed). So in the analyzer C we get resulting state $\ket{x}_{at}$  or $\ket{y}_{at}$  with equal probability.
Similar story with the atom in the initial state $(\ket{x}_{at}-i\ket{y}_{at})/\sqrt{2}$.

The first and the second modes of operation can therefore be distinguished, although they correspond to the same density matrix state of an atom. In the Mode 1 of operation of stage A detectors will be detecting only $\ket{y}_{at}$ atoms, while in the Mode 2 of operation - both $\ket{y}_{at}$ and $\ket{x}_{at}$ atoms.

As a conclusion, we have to admit, that the state of both atom and the photon should change in the event of non-interaction . It is important to note that scenarios of \cite{Ref8} \cite{Ref9} (where entanglement is appearing between particles and varies between no entanglement in the oppositely-pointing states like $\ket{y}_{ph}\ket{x}_{at}$ to maximal entanglement in cases when there's partial overlap as in state $\ket{y}_{ph}\ket{x+iy}_{at}$) also do not pass the criteria of this thought experiment and allow to distinguish indistinguishable ensembles. In order to change in such a way that keeps states of the same density matrix indistinguishable, a new state after non-interaction should be basis-of-description independent and symmetric. Moreover, to be logically consistent, both states should change during ANY event of non-interaction. This means that if before the interaction the state of the photon was $\ket{y}_{ph}$ 
and the state of the atom $\ket{x}_{at}$.Then after the non-interaction the state of both particle changes. 

We therefore suggest that as a result of process of non-interaction the probe and the object end up in the maximally entangled state like $(\ket{x}_{ph}\ket{y}_{at} - \ket{y}_{ph}\ket{x}_{at})/\sqrt{2}$ .  
So that before the measurement of the final state of one of the partners we don't know anything about the state of the other. At the same time, measurement of one is consistent with the state of the other. Moreover, this is the only two-particle entangled state that is the same in all the conjugate  bases and ensures that in any basis measurement reveals both particles pointing in the opposite direction.

In general, to test a possible scenario of non-interaction using the aforementioned thought experiment, following procedure is suggested. First, rules of interaction should be proposed. Second, use those rules for the device as it is described in the figure \ref{fig:filter}. Third, check whether those rules are consistent if the basis of filter and measurement (stages B and C) is chosen to be different. Fourth, repeat second and third point, but exchange object and the probe (stages A and B) and see if the same rules are consistent.
	
Another way in which a question similar to that investigated in this work might be put is the following. If we send a photon in the state $\ket{y}_{ph}$ 
onto the atom in the state $\ket{x}_{at}$, 
and after the event of non-interaction measure both of them in $\ket{\sigma+/-}$
basis – will we get a perfect negative correlation between the results for atom and the photon (will every measurement of $\ket{\sigma+}_{ph}$ coincide with $\ket{\sigma-}_{at}$ 
and other way around)? In light of the abovementioned thought experiment, there should be perfect correlation. 

In case there's  no correlation (which means that sometimes atom and the photon will be measured as aligned although they haven't interacted) we get into contradiction anain. Indeed, why the states that could have interacted never did so (assuming idealized (non)interaction)? 

There are two alternatives we've found that seem to fulfill the criteria of the experiment, but seem less appealing to us. First is both atom and photon would end up in absolutely random states after any event of non-interaction. Effectively, this would imply that the states of both atom and photon become mixed states described by identity density matrices or Bloch Spheres with a dot in the middle. This, however, seems to imply some information loss, which is doubtful. 

Second, there is a basis of preference onto which the states always collapse in the case of non-interaction. For example if we assume that atom-photon system collapses into $\sigma+ \sigma-$ basis always. However, the quantum rules for wavefunction collapse don't have a clause about a preferred basis of collapse.

The thought experiment proposed here provides conditions based on fundamental principles of quantum mechanics that every potential scenario of non-interaction has to fulfill. The fact that the roles of object and the probe can be switched between atom and photon (in stages A and B) ensures that non-interaction has to be symmetric both in terms of what happens with either of the particles and in terms of basis-independence of the final state.

If we were to state the most unintuitive, but at the same time indispensable and characteristic implication of our analysis, it would be the following. Imagine a single photon in the state $\ket{y}_{ph}$ probing a single atom in the state $\ket{x}_{at}$ so that the state of the system is $\ket{y}_{ph}\ket{x}_{at}$. As (non)interaction happens, they both get maximally entangled. So when we measure the photonic state in x/y basis after that, there is 50\% chance of seeing it in the state coinciding with its initial state, namely $\ket{y}_{ph}$ and 50\% chance of it being $\ket{x}_{ph}$. Because measurement of the state of one of the particles of the entangled pair is described by the unity density matrix, implying absolutely random result. Moreover, in case we measure the state of the photon to be $\ket{x}_{ph}$ (essentially its state has flipped), following the rules of measurement of entangled particles, we are sure to find the state of the atom to be  $\ket{y}_{at}$ and the state of the whole system now being $\ket{x}_{ph}\ket{y}_{at}$. Effectively, the state of both particles flips to the orthogonal one as a result of (non)interaction and a consecutive measurement.

It is important to note again that the atom and photon are idealized, so whenever they can interact, they will. And the event of non-interaction here is the one arising from misalignment of atomic and photonic states (bloch sphere representation), not from the event when photon has missed the atom, possibility of which is ruled out by assuming idealized atom and photon. But (!) the analysis with same results could be repeated also for interaction which is not perfect and all of the results will hold in that case too. Take, for instance, the analysis given in this section. Let's assume there is some probability that interaction between atom and photon does not occur (they ``fly by" each other), even if internal degrees of freedom were perfectly aligned. Imagine, this probability is 0.5. Statistics of interaction would change, of course. But this possibility of ``flying by" would alter the statistics in all cases in the same way. Namely, we will get random result in 50\% of the cases in any setup of the thought experiment. Essentially, introducing noise into this thought experiment does alter the results, but in the same way for all the cases. The only difference is the number of repetitions of the experiment that is needed to see the desired effect. So the analysis presented here doesn't lose generality while gaining considerably in clarity.

As has been mentioned earlier, the analysis presented here is not limited to the specific atom-photon non-interaction. It is rather an illustration of a more abstract case. Any two quantum objects that have two orthogonal states each, states, that are connected to each other through possible interaction (or non-interaction), can be candidates for similar analysis. One can perform essentially the same thought experiment with an electron and a positron passing through beamsplitters and having their spatial paths as the two possibly interacting states. 

\section{Analysis and Conclusions}
\label{conclusion}

It seems that the logic of ordinary IFM (IFM with a classical massive object) when the state of the object is assumed to be unalterable during the non-interaction isn't applicable in case when both the probe and the object are quantum objects. The state of neither of them can remain unchanged. This change of the state should be symmetric with respect to the exchange of roles of ''probe" and ''object" and satisfy conditions presented by the filtering device proposed in this work. Complete entanglement between the ``object" and the ``probe" in every event of non-interaction during the IFM satisfies both requirements.
The thought-experimental device presented in this analysis (fig. \ref{fig:filter}) is not only a vehicle for the derivation of results of this work. It also is a mechanism that allows to test ANY proposed explanation of non-interaction for consistency. All the explanations of non-interaction of two quantum objects that were available to authors did not pass the test of the device.

This analysis has interesting implications. First, entanglement does lie at the heart of non-interaction. Moreover, it is a defining feature of this process. Essentially, non-interaction is entanglement. 

Second, there is a way of creating entanglement of two possibly interacting two-level quantum systems. Third, we can leave atom in a desired superposition of ground states by the act of measurement of the photon and then sorting out and picking the relevant cases. 

An interesting further development of this work would be to look at the possibility of extending entanglement by introducing more events of non-interaction (for example chaining multiple atoms using the same photon). Or entangling already entangled probe-object pairs, creating four- and more-partite entanglement. In this case some elements of the chain would get entangled with each other without ever having to even possibly interact (or non-interact).  
\newpage
\appendix
\section{Formal proof of aligning Bloch spheres representing interacting particles}

In words this proof shows that:

given two particles with two degrees of freedom each and those DoFs being coupled to each other 

for every superposition of those DoFs in one particle there is a corresponding superposition in the DoFs of the other particle that will result in an event of absorbtion in 100\% of the cases.

\bigskip

We start with postulating what we know. \smallskip

$\exists$ an atomic state $\ket{?}_{at}$ s.t. given an idealized atom, it absorbs photon in state $\ket{x}_{ph}$ in 100\% of the cases. Let us call such a state $\ket{x}_{at}$

This state $\ket{x}_{at}$ absorbs $\ket{y}_{ph}$ in 0\% of the cases. And $\bra{x}_{ph}\ket{y}_{ph}=0$

Then

$\exists$ $U_{x\rightarrow y}^{at}$ s.t. $U_{x\rightarrow y}^{at}\ket{x}_{at}$ absorbs $\ket{y}_{ph}$ in 100\% of the cases. Let's call $U_{x\rightarrow y}^{at}\ket{x}_{at} = \ket{y}_{at}$ 

State $\ket{x}_{at}$ absorbs $(\ket{x}_{ph}+i\ket{y}_{ph})/\sqrt{2} = \ket{\sigma+}_{ph}$ in 50\% of cases.

$\exists$ $U_{x\rightarrow \sigma+}^{at}$ s.t.  $(U_{x\rightarrow \sigma+}^{at})^2\ket{x}_{at} = U_{x\rightarrow y}^{at}\ket{x}_{at} = \ket{y}_{at}$ and $U_{x\rightarrow \sigma+}^{at}\ket{x}_{at}$ absorbs $\ket{\sigma+}_{ph}$ in 100\% of the cases.

Same for the state $\ket{\sigma-}_{at}$ that absorbs in 100\% of the cases the state $\ket{\sigma-}_{ph}$

\begin{acknowledgements}
We would like to express our gratitude to the following people and organizations who made this work possible. Rin who provided inspiration for this work. Vyacheslavs Kashcheyevs for his comments, criticism and all the help. MIT OCW, Leonard Susskind 
%and Complexity Explorer
for providing knowledge. Douglas Hofstadter and Mark Strand. Carl Vandivier, TMBCC and Morgan Book for great discussions and atmosphere in Bloomington. The support of our families has been constant and invaluable.
\end{acknowledgements}


\begin{thebibliography}{}
\bibitem{Jacobs2014}
K. Jacobs, Quantum Measurement Theory and its Applications, Cambridge University Press (2014)
\bibitem{Ren1953}
M. Renninger, (1953) Z. Phys, 136 p 251
\bibitem{EV1993}
A. C. Elitzur, and L. Vaidman, Found. Phys., 23, 987 (1993)
\bibitem{Kwiat1995}
P. Kwiat, H. Weinfurter, T. Herzog, A. Zeilinger, and M. Kasevich, Phys. Rev. Lett.,74, 4763 (1995).
\bibitem{Ref3}
P. Kwiat, H. Weinfurter, A. Zeilinger, 673-674, Coherence and Quantum Optics VII, Plenum Press, New York (1996)
\bibitem{Ref4}
L.Hardy, Phys.Rev.Lett., 68, 2981-2984 (1992)
\bibitem{Ref10}
Y. Aharonov, A. Botero, S. Popescu, B. Reznik, J. Tollaksen, Phys.Lett.A, 301, 130-138 (2002)
\bibitem{Ref11}
W. Irvine, J. Hodelin, C. Simon, and D. Bouwmeester, Phys.Rev.Lett., 95, 030401 (2005)
\bibitem{Ref12}
J.S. Lundeen, A.M. Steinberg, Phys.Rev.Lett., 102, 020404 (2009)
\bibitem{Ref13}
K. Yokota, T. Yamamoto, M. Koashi, and N. Imoto, New J. Phys, 11 (2009)
\bibitem{Ref5} 
H. Azuma, Phys.Rev.A, 70, 012318 (2004)
\bibitem{Ref6}
Y. P. Huang, and M. G. Moore, Phys.Rev.A, 77, 062332 (2008)
\bibitem{Ref14}
O. Hosten, M. Rakher, J. Barreiro, N. Peters, and P. Kwiat, Nature, 439, 949-952 (2006)
\bibitem{Ref15}
L. Vaidman, Phys.Rev.Lett, 98, 160403 (2007)
\bibitem{Ref16}
F. Kong, C. Ju, P. Huang, P. Wang, X. Kong, F. Shi, L. Jiang, and J. Du, Phys.Rev.Lett., 115, 080501 (2015)
\bibitem{Ref17} 
H. Salih, Z-H. Li, M. Al-Amri, and S. Zubairy, Phys.Rev.Lett., 110, 170502 (2013)
\bibitem{Ref18}
Y. Cao, Y-H. Li, Z. Cao, J. Yin, Y-A. Chen, X. Ma, C-Z. Peng, J-W. Pan, Lasers and Electro-Optics (CLEO) Conference on (2014)
\bibitem{Ref19}
T-G. Noh, Phys.Rev.Lett, 103, 230501 (2009)
\bibitem{Ref20}
Y. Aharonov, E. Cohen, A.C Elitzur, L. Smolin, Found. Phys. 48, 1-16 (2018)
\bibitem{Ref7}
S. Potting, E.S. Lee, W. Schmitt, I. Rumyantsev, B. Mohring, P. Meystre, Phys.Rev.A, 62, 060101(R) (2000) 
\bibitem{Ref8}
X. Zhou, Z. Zhou, G. Guo, and M. Feldman, Phys.Rev.A, 64, 020101 (2001)
\bibitem{Ref9}
R. Angelo, Found.Phys., 39, 109-119, (2009)
\bibitem{Ref21}
M. O. Scully, B.-G. Englert, H. Walther, Nature, 351(6322), 111–116 (1991)
\bibitem{Ref22}
L.C. Ryff, P.H. Souto Ribeiro, Phys. Rev. A 63, 023801, (2001)
\bibitem{Brag80}
V. B. Braginsky, Y. I. Vorontsov, and K. S. Thorn, Science 209 547 (1980)
\bibitem{Putz2010}
M.V.Putz, Int. J. Mol. Sci. 2010 Oct 22;11(10):4124-39
\bibitem{Putz2010bond}
M.V.Putz, Int. J. Mol. Sci. 2010, 11(11)
\bibitem{Ref23}
L. Vaidman, Found.Phys., 33, 491-510, (2003)
\bibitem{Neu2018} John von Neumann, Mathematical Foundations of Quantum Mechanics, Princeton University Press; New, Translation edition (February 27, 2018).
\bibitem{Blum}
K. Blum, Density Matrix Theory and Applications, Springer-Verlag Berlin Heidelberg (2012)


\end{thebibliography}


\end{document}